# EFFECTS OF NEUTRON-PROTON INTERACTION IN DOUBLY ODD DEFORMED NUCLEI*

N. Itaco, A. Covello, and A. Gargano

Dipartimento di Scienze Fisiche, Università di Napoli Federico II
and Istituto Nazionale di Fisica Nucleare
Complesso Universitario di Monte S. Angelo, Via Cintia, 80126 Napoli, Italy

We have investigated the effects of the neutron-proton interaction in several doubly odd deformed nuclei within the framework of the particle-rotor model. In this paper, we show some selected results of our study which evidence the importance of the tensor-force effects.

## 1. Introduction

As is well known, the two most important effects associated with the residual interaction between the odd neutron and the odd proton in doubly odd deformed nuclei are the Gallagher-Moszkowski (GM) splitting [1] and the Newby (N) shift [2]. In addition relevant information may be provided by the study of the odd-even staggering in $K \neq 0$ bands [3, 4]. In fact, this effect may be traced to direct Coriolis coupling of $K \neq 0$ bands with one or more N-shifted $K = 0$ bands.

As early as some thirty years ago, several efforts [5-7] were made to obtain detailed information on the effective neutron-proton interaction in the rare-earth region from the empirical values of GM splittings and N shifts. In this context, the importance of the tensor-force effects for the N shifts was pointed out [7]. In the following years, however, this problem received very little attention, a simple spin-dependent $\delta$ force being adopted in most calculations [8].

In this situation, we found it interesting to carefully investigate the effects of the neutron-proton interaction in doubly odd deformed nuclei focusing attention on the role of the tensor force. To this end, we have studied [4, 9, 10] several doubly odd isotopes in the rare-earth region within the framework of the particle-rotor model. Here, we present only some selected results concerning N-shifted $K = 0$ bands. As we shall see in Sec. 3, the experimental data are very well reproduced when using the central plus tensor force proposed by Boisson *et al.* [7].

## 2. Outline of the model and calculations

We assume that the unpaired neutron and proton are strongly coupled to an axially symmetric core and interact through an effective interaction. The total Hamiltonian is then written as

$$H = H_0 + H_{\mathrm{RPC}} + H_{\mathrm{ppc}} + V_{np}. \tag{1}$$

The term $H_0$ includes the rotational energy of the whole system, the deformed, axially symmetric field for the neutron and proton, and the intrinsic contribution from the rotational degrees of freedom. It reads

$$H_0 = \frac{\hbar^2}{2\mathcal{J}}(\boldsymbol{I}^2 - I_3^2) + H_n + H_p + \frac{\hbar^2}{2\mathcal{J}}[(\boldsymbol{j}_n^2 - j_{n3}^2) + (\boldsymbol{j}_p^2 - j_{p3}^2)]. \tag{2}$$

The two terms $H_{\mathrm{RPC}}$ and $H_{\mathrm{ppc}}$ in Eq. (1) stand for the Coriolis coupling and the coupling of particle degrees of freedom through the rotational motion, respectively. Their explicit expressions are

$$H_{\mathrm{RPC}} = -\frac{\hbar^2}{2\mathcal{J}}(I^+J^- + I^-J^+), \tag{3}$$

$$H_{\mathrm{ppc}} = \frac{\hbar^2}{2\mathcal{J}}(j_n^+ j_p^- + j_n^- j_p^+). \tag{4}$$

The effective neutron-proton interaction has the general form

$$V_{np} = V(r)[u_0 + u_1\boldsymbol{\sigma}_p\cdot\boldsymbol{\sigma}_n + u_2 P_M + u_3 P_M \boldsymbol{\sigma}_p\cdot\boldsymbol{\sigma}_n + V_T S_{12} + V_{TM} P_M S_{12}], \tag{5}$$

with standard notation [7]. In our calculations we have used a finite-range force with a radial dependence $V(r)$ of the Gaussian form as well as a zero-range force,

$$V_{np}^{\delta} = \delta(r)[v_0 + v_1\boldsymbol{\sigma}_p\cdot\boldsymbol{\sigma}_n]. \tag{6}$$

We have used the standard Nilsson potential [11] to generate the single-particle Hamiltonians $H_n$ and $H_p$. The parameters $\mu$ and $\kappa$ have been fixed by using the mass-dependent formulas of Ref. [12]. The deformation parameter $\beta_2$ has been deduced for each doubly odd isotope from the neighboring even-even nucleus while the single-particle energies for the odd proton and the odd neutron and the rotational parameter $\frac{\hbar^2}{2\mathcal{J}}$ have been derived from the experimental spectra of the two neighboring odd-mass nuclei.

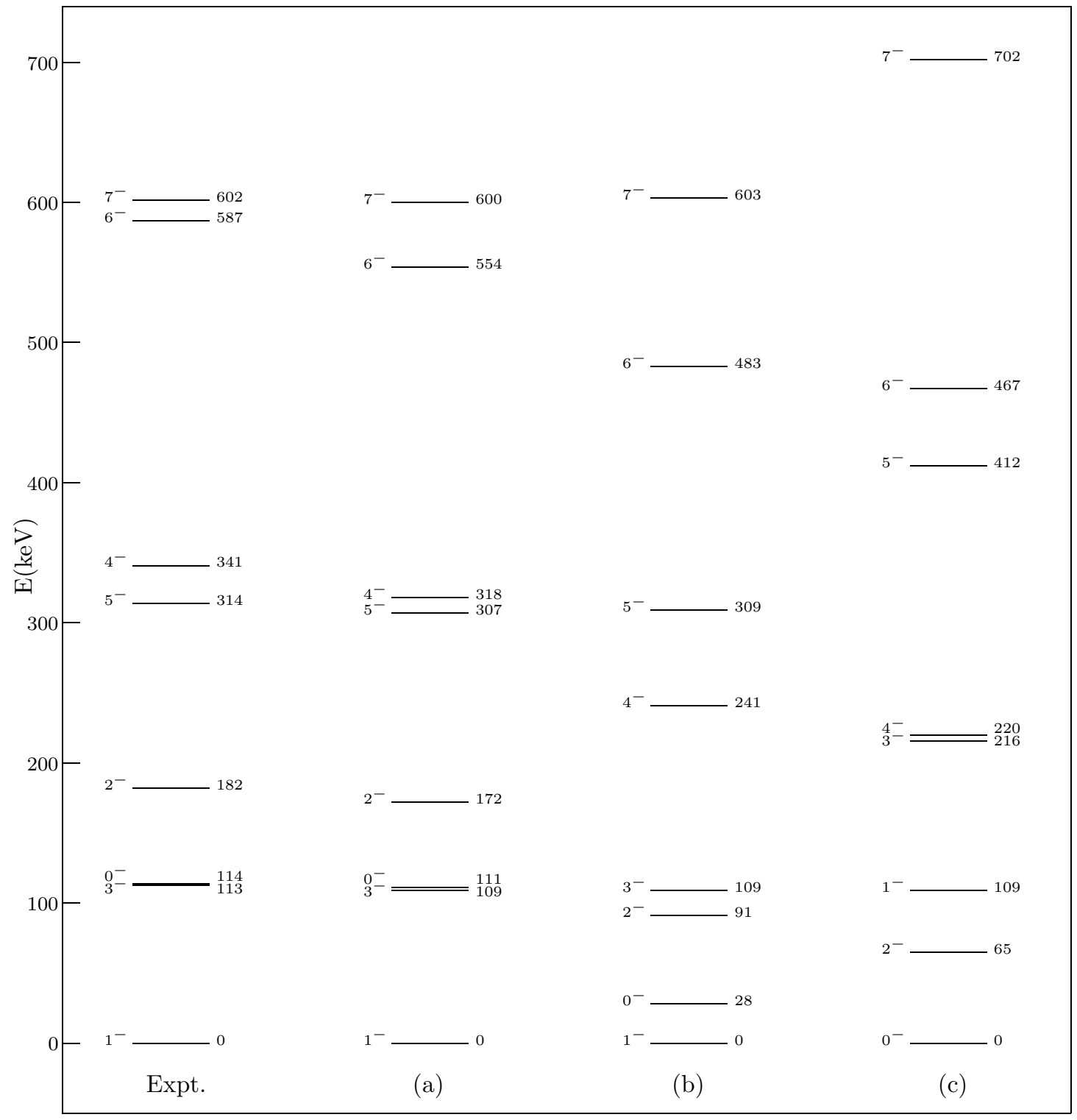

Fig. 1. Experimental and calculated spectra of the lowest $K^\pi = 0^-$ band in $^{176}$Lu. The theoretical spectra have been obtained by using (a) a central plus tensor force with a Gaussian radial shape, (b) a Gaussian central force, and (c) a $\delta$ force.

To explore the role of the tensor force, we have performed two different calculations with the finite-range interaction, with and without the tensor terms, respectively. In both cases for the parameters of the interaction we have used the values determined by Boisson *et al.* [7] in their study of doubly odd nuclei in the rare-earth region. As regards the $\delta$ force, we have used for the strength of the spin-spin term $v_1$ the value -0.20 MeV, which leads to the lowest possible disagreement between theory and experiment. More details about the choice of the parameters of the potentials as well as their values can be found in Ref. 4.

## 3. Results and comparison with experiment

In this section, by way of illustration, we present some selected results of our study. More precisely, we consider three $K = 0$ bands corresponding to different intrinsic configuration identified in $^{176}$Lu, $^{182}$Ta and $^{188}$Re, re-spectively. As regards the results concerning the odd-even staggering effect, they may be found in Refs. [4,9,10].

In Fig. 1 we compare the experimental spectrum of the $K^\pi = 0^- p\frac{7}{2}[404]$-$n\frac{7}{2}[514]$ band in $^{176}$Lu [13] with the spectra obtained by using the Gaussian force, with and without tensor terms, and the $\delta$ force. We see that the right level order and an impressive agreement with experiment is obtained when using the Gaussian force with tensor terms. The experimental N shift (70 keV) is very well reproduced (66 keV), and the largest discrepancy in the excitation energies is only about 30 keV.

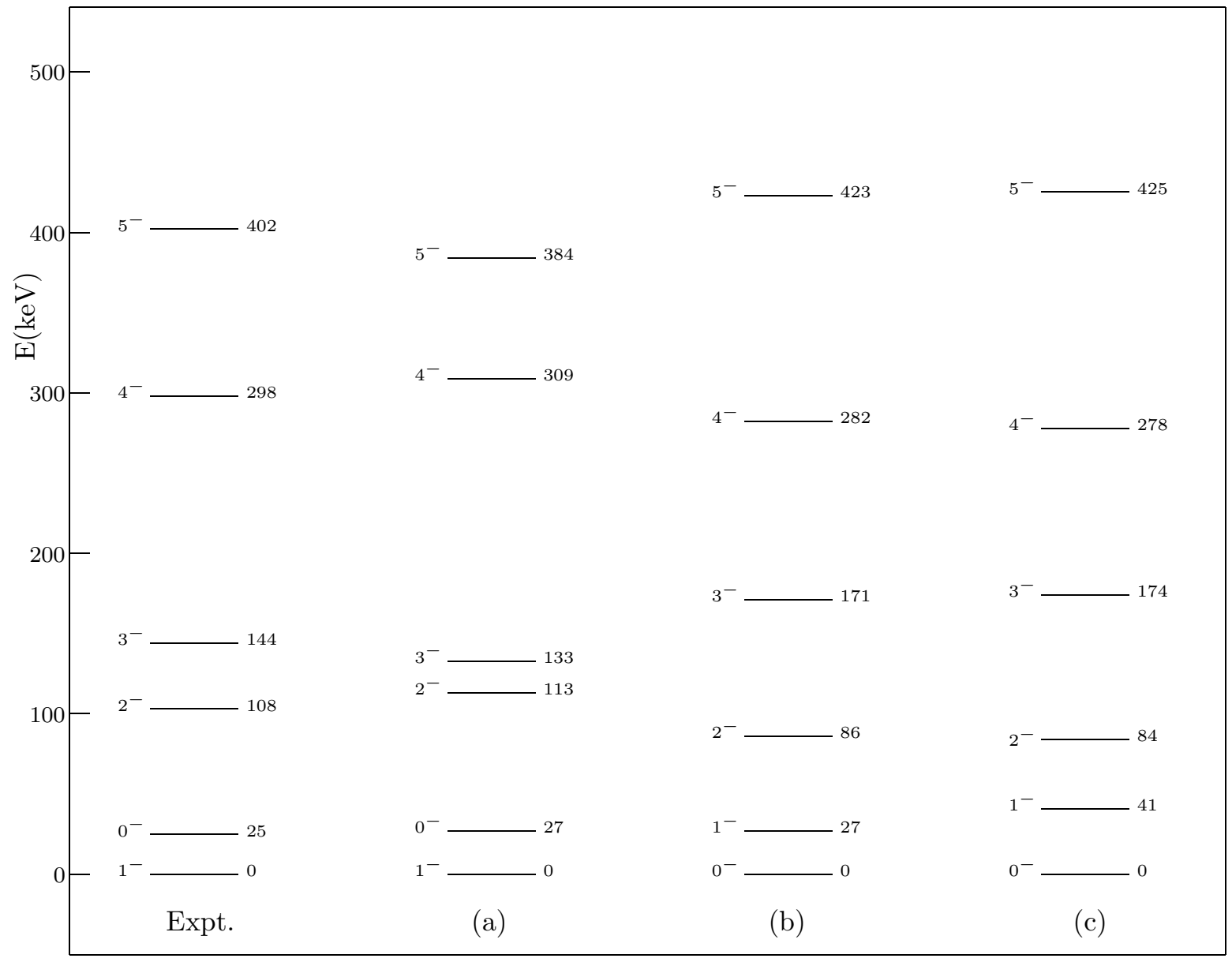

Fig. 2. Same as Fig. 1, but for the lowest $K^\pi = 0^-$ band in $^{182}$Ta.

The importance of the tensor force is confirmed, as shown in Fig. 2, when studying the $K^\pi = 0^- p\frac{7}{2}[404]n\frac{7}{2}[503]$ band in $^{182}$Ta. In this case, the values of the calculated N shift [10] are 28, 1, and -6 keV, using the Gaussian force with tensor terms, the Gaussian central force, and the $\delta$ force, respectively. These values are to be compared with the experimental value of 26 keV [13].

Finally, the three calculated spectra for the $K^\pi = 0^+ p\frac{9}{2}[514]n\frac{9}{2}[505]$ band in $^{188}$Re are compared with the experimental one [13] in Fig. 3. Also in this case only the Gaussian force with tensor terms is able to give a satisfactory reproduction of the experimental spectrum. The calculated N-shift values are -62, -3, and -1 keV in case (a), (b), and (c) respectively, to be compared with the experimental value of -54 keV.

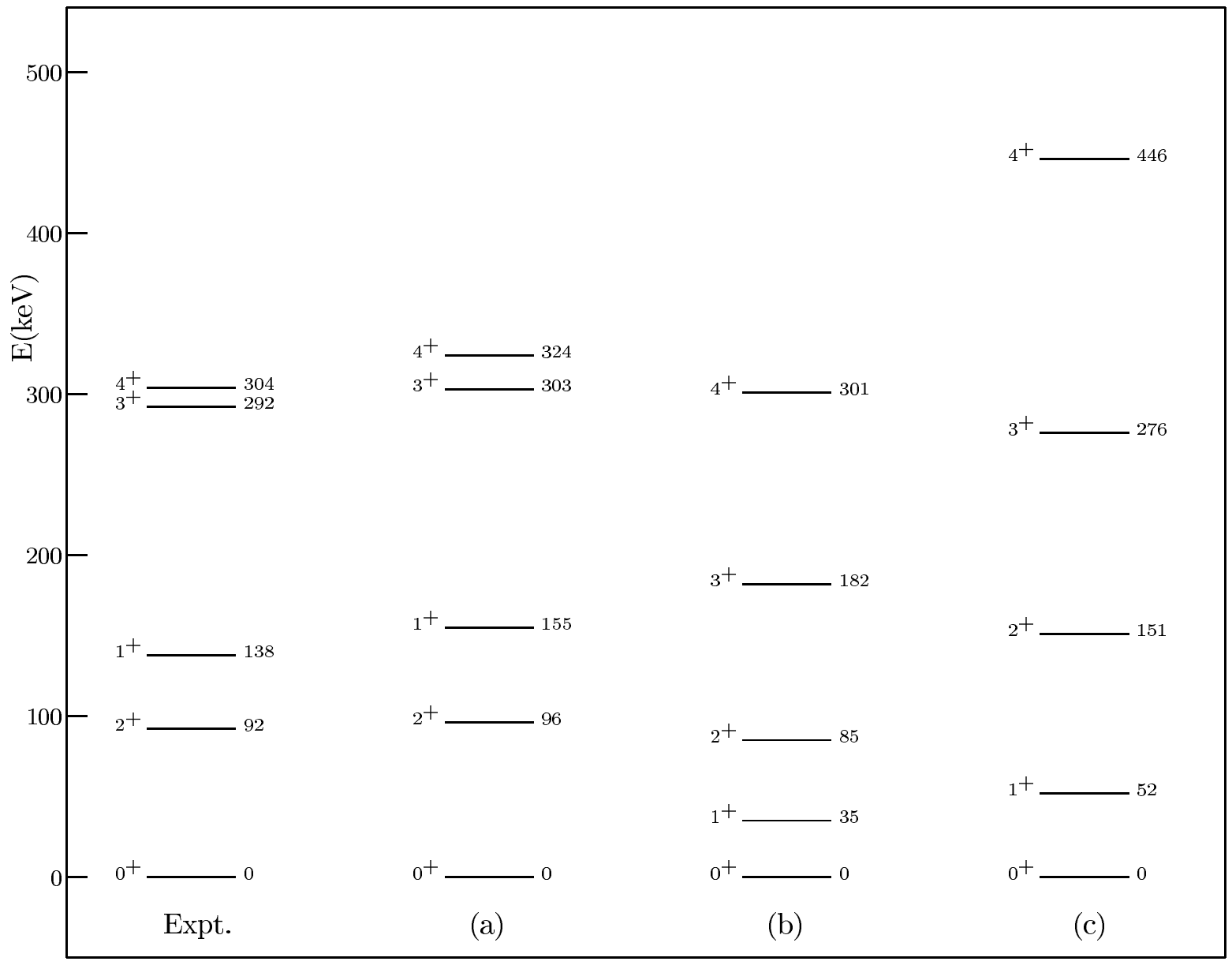

Fig. 3. Same as Fig. 1, but for the lowest $K^{\pi} = 0^{+}$ band in $^{188}$Re.

## 4. Summary

In this paper, we have presented some selected results of a particle-rotor model study of doubly odd deformed nuclei in the rare-earth region. This work has been aimed at obtaining detailed information on the effective neutron-proton interaction, with particular attention focused on the role of the tensor force, which has been neglected in most of the existing studies to date. The results of our calculations evidence the importance of this force for the description of some relevant aspects of the structure of doubly odd deformed nuclei.